\documentclass{article}
\usepackage{spconf,amsmath,graphicx}

\usepackage{booktabs} 
\usepackage{xspace}
\usepackage{multirow}
\usepackage{colortbl}
\usepackage[table,xcdraw]{xcolor}
\usepackage{tabularx}
\usepackage{todonotes}
\usepackage{graphicx}
\usepackage[hidelinks]{hyperref}

\usepackage{cite}

\usepackage{subcaption}

\definecolor{mycolor}{HTML}{FF6600}
\definecolor{indiagreen}{HTML}{138808}
\definecolor{papaya}{HTML}{EE892F}
\definecolor{mygreen}{HTML}{008000}
\definecolor{mypurple}{HTML}{9966CC}
\definecolor{myblue}{HTML}{5D8AA8}
\definecolor{mypink}{HTML}{EC008C}

\title{Understanding the strengths and weaknesses of SSL models \\for audio deepfake model attribution}
\name{Gabriel Pîrlogeanu$^1$, Adriana Stan$^{1,2}$, Horia Cucu$^1$
}
\address{$^1$Speech and Dialogue Research Laboratory, POLITEHNICA Bucharest, Romania\\
$^2$Communications Department, Technical University of Cluj-Napoca, Romania\\
\small \texttt{gabriel.pirlogeanu@upb.ro, adriana.stan@com.utcluj.ro, horia.cucu@upb.ro}
}

\newcommand{\bert}{\texttt{w2v-bert-2.0}\xspace}
\newcommand{\xls}{\texttt{wav2vec2-xls-r-2b}\xspace}

\begin{document}

\maketitle
\begin{abstract}

Audio deepfake model attribution aims to mitigate the misuse of synthetic speech by identifying the source model responsible for generating a given audio sample, enabling accountability and informing vendors. The task is challenging, but self-supervised learning (SSL)-derived acoustic features have demonstrated state-of-the-art attribution capabilities, yet the underlying factors driving their success and the limits of their discriminative power remain unclear. In this paper, we systematically investigate how SSL-derived features capture architectural signatures in audio deepfakes. By controlling multiple dimensions of the audio generation process we reveal how subtle perturbations in model checkpoints, text prompts, vocoders, or speaker identity influence attribution. Our results provide new insights into the robustness, biases, and limitations of SSL-based deepfake attribution, highlighting both its strengths and vulnerabilities in realistic scenarios.

\end{abstract}

\begin{keywords}
audio deepfakes, model attribution, checkpoints, SSL, anti-spoofing, source tracing.
\end{keywords}

\section{Introduction}
\label{sec:intro}

Deepfakes have emerged as a critical societal challenge, with synthetic audio enabling high-stakes fraud, impersonation, and disinformation. While most research focuses on binary detection (distinguishing fake from real), this is insufficient for forensic accountability. Effective countermeasures require model attribution, i.e., identifying the specific system or vendor responsible for generating a deepfake. 
Consequently, attribution methods must remain robust under conditions where models are retrained, fine-tuned, or otherwise updated, ensuring reliable identification in realistic, evolving threat scenarios.
We further clarify that checkpoint attribution identifies a model’s fixed weights, while implementation attribution or source tracing identifies the codebase used for a specific audio generation architecture.

Several recent studies have addressed the problem of audio deepfake attribution. Müller et al.~\cite{muller22b_interspeech} use handcrafted and neural features for clustering deepfake samples by their generating attack system in ASVspoof 2019~\cite{asv19}. 
Zhu et al.~\cite{zhusource22} apply ResNets for multi-task classification of generators, conversion methods, and speakers. Klein et al.~\cite{klein2024source}, which combines various front-end architectures, such as ResNet, self-supervised learning, and processes them through AASIST~\cite{Jung2021AASIST} or the Whisper encoder~\cite{radford2022robustspeechrecognitionlargescale}. The AASIST architecture is also applied in~\cite{xie24_interspeech} on top of SSL features for in- and out-of-domain classification with the ADD2023 dataset.
Falez et al.~\cite{falez25_interspeech} perform fine-grained source tracing, finding vocoder attribution especially challenging. 

All prior work showed that deepfake model attribution becomes harder as the variability of generation models and parameters is extended. Within this landscape, SSL models have shown some of the most consistent results. In our prior work~\cite{stan_interspeech25} we used two SSL models
combined with a simple kNN classifier to obtain a strong checkpoint attribution and out-of-domain detection. 
However, the work relied on multiple datasets, and could not control for factors such as: implementation, speaker, prompts, or vocoder. As a result, subtle differences between implementations led to reduced architecture-level attribution accuracy.

In this study, we investigate how such SSL-derived features behave under controlled variations in the speech generation pipeline. We explicitly isolate implementation, training protocol, prompts, vocoder, and speaker, retrain several speech synthesis architectures from scratch, and evaluate the official pretrained checkpoints. 

The main \textbf{contributions} of this paper can be summarised as follows: (i) we retrain four speech generation systems from scratch using controlled conditions for data and compute procedures; (ii) we analyse how model attribution is affected by minor perturbations in a speech generation systems' output; (iii) we show that different SSL models have non-overlapping weak points, but still remain a powerful tool in identifying the source of an audio deepfake. 

\section{Methodology}
\label{sec:meth}

\subsection{Speech generation models}

\textbf{Training data.} To control for speaker variability, we use the LJSpeech dataset~\cite{ljspeech17}, which contains 13,100 audio samples from a single female speaker reading different text prompts, totalling 24 hours of recordings. For model training, we adopt the standard split of 12,500 audio samples. For model attribution evaluation, we select 600 text prompts for the validation and test sets, which are then used to generate synthetic audio samples across all speech generation architectures. 
We also conduct light speaker adaptation experiments using 500 randomly selected audio samples from HiFi-TTS speaker \textit{9136} (matching LJSpeech gender) and generate evaluation samples with the same set of 600 text prompts.

\textbf{Architectures and implementations.} We select four of the most commonly used text-to-speech architectures and their official implementations: FastPitch
~\cite{fastpitch} is a fully non-autoregressive text-to-speech model designed for fast, high-quality speech synthesis with explicit prosody control.
VITS
~\cite{vits} is a neural TTS framework that integrates text-to-spectrogram conversion and waveform generation into a single probabilistic model, enabling high-quality and efficient speech synthesis without a separate vocoder.  
Grad-TTS
~\cite{grad} is a diffusion-based TTS model that leverages score-based generative modeling to synthesize natural and expressive speech by progressively refining samples from noise toward a target mel-spectrogram.  
Finally, Matcha-TTS
~\cite{matcha} is a diffusion-based text-to-speech framework that improves upon prior score-based models by introducing efficient training and sampling strategies, enabling natural-sounding speech generation with reduced computational cost.


\textbf{Vocoders.} For Grad-TTS and Matcha-TTS, we employ the pretrained HiFi-GAN~\cite{hifigan} vocoders linked in their respective repositories.\footnote{\url{https://drive.google.com/file/d/15AeZO2Zo4NBl7PG8oGgfQk0J1PpjaOgI/view}, \\ \url{https://drive.google.com/file/d/18TNnHbr4IlduAWdLrKcZrqmbfPOed1pS/view}} For FastPitch we use a HiFi-GAN checkpoint from NGC.\footnote{\url{https://catalog.ngc.nvidia.com/orgs/nvidia/teams/dle/models/hifigan__pyt_ckpt_mode-finetune_ds-ljs22khz}} VITS integrates the HiFi-GAN generator directly in its encoder--decoder architecture.

\textbf{Training protocol and checkpoints.} As any vendor exposes a core model from which voice cloning or adaptation functionalities are derived, we first use the official pretrained models on LJSpeech from each architecture. 
Each architecture was trained from scratch for $500$k iterations on the LJSpeech data. We retain nine checkpoints at early ($50$k, $75$k, $100$k), mid ($250$k, $275$k, $300$k) and late ($450$k, $475$k, $500$k) training stages. 
The speaker adaptation process is performed by finetuning the pretrained LJSpeech models on $500$ samples from speaker $9136$ for $10$k iterations. 
To investigate potential architecture-level inherent biases that may manifest in trained models, we preserve the initializations (``zero-init'') from three random seeds for each architecture. 

Training was carried out under a unified compute setup: a single NVIDIA T4 GPU with a batch size of~$32$. All other parameters were fixed at their repositories' default settings.

\subsection{Audio deepfakes attribution system}
\vspace{-.2cm}
We follow the setup from Stan et al.~\cite{stan_interspeech25} and maintain the lightweight attribution system based on k-Nearest Neighbours (kNN) with time-domain average pooled SSL features. This ensures that we can prioritize the analysis of SSL-derived features themselves, rather than the classification capabilities of a complex DNN, for example. To understand potentially different SSL behaviours, we retain both models.
To reiterate, each audio deepfake sample is represented by either a $1920$-dimensional feature vector derived from \xls or a $1024$-dimensional vector derived from \bert. The vectors serve as support and query in the kNN-based attribution.

\vspace{-.2cm}
\subsection{Evaluation protocol}
\vspace{-.2cm}

From the 600 text prompts, 420 samples (70\%) form the support set, while 60 (10\%) and 120 (20\%) samples are used for validation and query, respectively. Classes are balanced without weighting. Experiments are in-domain (ID) if support and query set samples come from the same checkpoints, and out-of-domain (OOD) otherwise. Deviations from this data partitioning are explicitly noted in the results section.
Performance is evaluated using the \textbf{macro F1-score}. We distinguish between \textit{checkpoint attribution}—identifying specific model weight states (checkpoints)—and \textit{architecture attribution}—grouping all checkpoints from the same generator into one of four architecture classes, a comparatively simpler task. All trained models and generated audio samples are made freely available upon request.


\vspace{-.2cm}
\section{Results}
\label{sec:res}

The main objective of this study is to examine perturbation factors affecting SSL-based deepfake attribution by generating multiple data splits to capture potential biases in the model's discriminative power.

\begin{figure}[t!]
    \centering
    \hspace*{-0.5cm}
        \includegraphics[width=1.1\linewidth]{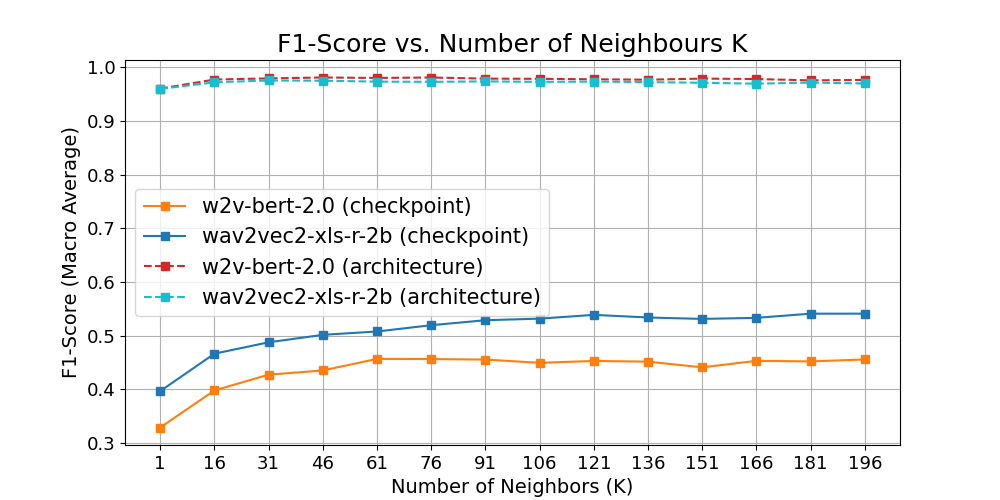}
       
     \caption{F1-scores for architecture and checkpoint attribution of the validation set as a function of $k$--number of neighbours.}
    \label{fig:ksel}
    \vspace{-.2cm}
\end{figure}

\textbf{Number of neighbours selection ($k$).} Following~\cite{stan_interspeech25}, we retain the best-performing layers (layer~8 and layer~4 for \xls and \bert, respectively), but recompute the optimal number of neighbours $k$. Attribution is performed in-domain, with disjoint text prompts across support, validation, and query sets. Figure~\ref{fig:ksel} shows validation performance as a function of $k$.
For architecture attribution, the SSL models perform similarly, while for checkpoint attribution \xls achieves an average F1-score improvement of $0.1$ over \bert. Both models plateau around $k=91$. To balance performance with practical constraints on sample availability, we fix $k=56$ for subsequent evaluations.


\textbf{Baseline attribution system.} The baseline results in this study refer to an in-domain attribution over the query set of samples. We use the 9 trained, and 1 pretrained checkpoint from each of the four architectures, yielding a total number of 40 attribution classes. 
As shown in Table~\ref{tbl:baseline} (row~1), both models achieve high accuracy at architecture-level ($F1=0.98$), consistent with~\cite{stan_interspeech25}, while checkpoint attribution is notably weaker in this case ($F1\approx0.50$)--an expected result when using several checkpoints from the same model implementation. Although seemingly a low F1-score, a random chance classifier with 40 classes would only achieve an $F1=0.025$.

\begin{figure}[t!]
    \centering
     \includegraphics[width=0.75\linewidth]{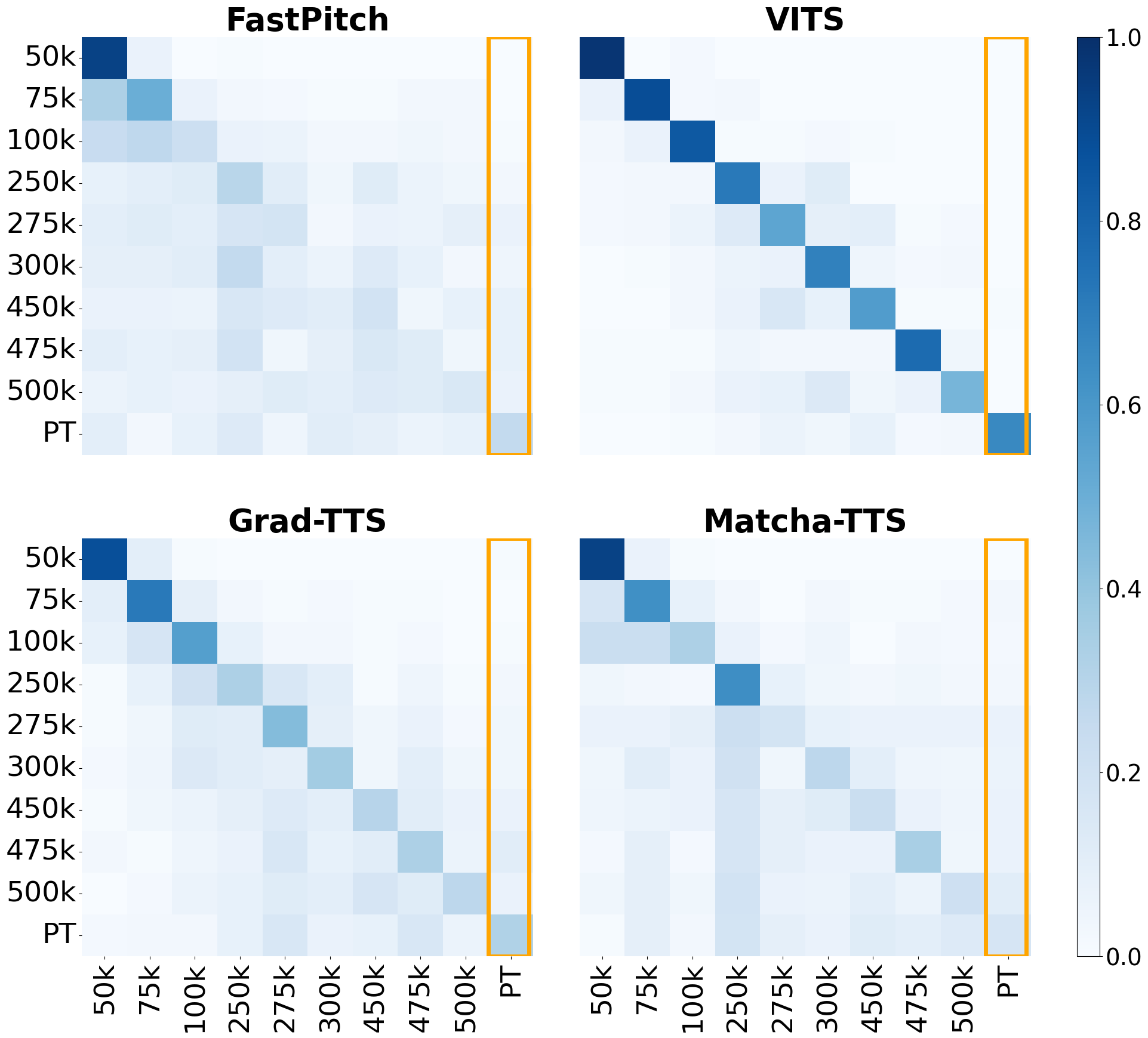}
        \caption{Baseline confusion matrices for each architecture. 
        The orange highlight represents the pretrained (PT) model.
        }
        \label{fig:baseline}
        \vspace{-.5cm}
\end{figure}

To analyse this effect, Figure~\ref{fig:baseline} presents checkpoint-level confusion matrices using \bert features. Given minimal cross-architecture attributions, a simplified view suffices, though all architectures appear in both support and query sets. Similar trends were observed with \xls. Overall, VITS checkpoints are more easily distinguished irrespective of the training stage (darker shades for diagonal values), whereas checkpoints from the other three architectures become less separable as training converges, since attribution should reflect architecture rather than training stage.
 Additional insights could potentially be drawn about the architectures themselves: VITS may require longer training with larger output changes per iteration, whereas FastPitch appears easier to train, exhibiting smaller output variations after 250k iterations. Also VITS' internal vocoder may potentially cause greater output variability across the training stages compared to the other architectures which use a fixed, external vocoder.

\textbf{OOD attribution}. Previous results relied on balanced splits with identical classes in support and query sets. Real-world scenarios often provide only a few core-model samples for attribution of unseen audio deepfakes. To simulate this, we created new splits (Table~\ref{tbl:baseline}, rows~2–4) with disjoint checkpoints in support and query sets. In such cases, only architecture-level attribution is applicable. Using only the pretrained model to attribute earlier checkpoints of a speech generator (row~2) slightly reduces performance, particularly for \xls. Increasing support-set variation (rows~3–4) mitigates these differences. 
Inspection of the confusion matrices reveals an unexpected but desirable pattern. Rather than being attributed primarily to trained models in the support set, many query samples are also attributed to the pretrained models. This indicates that the method emphasizes architecture-level cues over checkpoint-specific similarities. Cross-architecture attribution remains minimal.


\begin{table*}[t!]
    \centering
    \footnotesize
    \newcommand{\ii}[1]{{\footnotesize \color{gray} #1}}
    \setlength{\tabcolsep}{4pt}
    \caption{F1-scores for the checkpoint and architecture attribution systems across different sets of support and query sets. PT stands for the pretrained model. The checkpoints are taken from each architecture, so there are actually 4 times the number reported in the table. We also average the results across three random splits and report the stats.}
    \begin{tabular}{rlllll|cc|cc}
        \toprule
             &\textbf{Test}&\textbf{Support} & \textbf{Query}&   \textbf{Prompts} & \textbf{Vocoder} &   \multicolumn{2}{c}{\textbf{\xls}}  &\multicolumn{2}{c}{\textbf{\bert}}\\ 
             &\textbf{condition}& \textbf{ckpts} & \textbf{ckpts} &\textbf{ split }& \textbf{type}& \textbf{Checkpoint} & \textbf{Architecture} & \textbf{Checkpoint} & \textbf{Architecture} \\
        \midrule
        \ii{1} &Baseline& PT + 9 & PT + 9 & disjoint & default & 0.519±0.003 & 0.976±0.003 & 0.450±0.004 & 0.983±0.004\\ \midrule

        \ii{2} &OOD    & PT & 9 & disjoint & default & \ii{n/a} &  0.863±0.006 & \ii{n/a}& 0.923±0.007\\ 
        \ii{3} &       & PT + 50k+250k+450k  & others  & disjoint & default    & \ii{n/a}& 0.951±0.003 & \ii{n/a} & 0.971±0.006\\ 
        \ii{4}&       & PT + 5 random     & others  & disjoint & default    & \ii{n/a}  & 0.960±0.002 & \ii{n/a}&  0.975±0.006\\ \midrule

        \ii{5} &Prompts& PT + 9 & PT + 9 & mixed    & default & 0.432±0.003 & 0.973±0.001 & 0.367±0.000 & 0.978±0.001 \\ \midrule

        \ii{6} &Vocoder& PT + 9 & PT + 9 & disjoint & same     & 0.504±0.004 &  0.941±0.002&  0.436±0.004 & 0.943±0.004\\
        \ii{7} &       & PT + 9 & PT + 9 & disjoint & mixed    & \ii{n/a} &  0.634±0.006& \ii{n/a} & 0.551±0.003\\ \midrule

        \ii{8} &Speaker& PT + finetuned & PT + finetuned  & disjoint & default    & 0.994±0.001 & 0.994±0.001&  0.989±0.004 &  0.989±0.004\\ 
        \ii{9} &       & PT & finetuned     & disjoint & default    & \ii{n/a}  &  0.361±0.004 &  \ii{n/a}&  0.657±0.017\\ 
        \ii{10} &       & PT + 9& finetuned  & disjoint & default    & \ii{n/a}  &  0.342±0.001 &  \ii{n/a}&  0.643±0.001\\\midrule

        \ii{11} & Zero-init & Zero-init & Zero-init  & disjoint & default    & 0.874±0.001 &  1.000±0.000& 0.859±0.010 & 1.00±0.000\\ 
        \ii{12} &           & Zero-init & PT + 9  & disjoint & default    & \ii{n/a} &  0.100±0.00 & \ii{n/a} & 0.100±0.00\\ 
                
        
        \bottomrule
    \end{tabular}
    \label{tbl:baseline}
    \vspace{-.2cm}
\end{table*}

\textbf{Dependency on text prompts.}
SSL models primarily encode linguistic information, as their primary downstream task is speech recognition. We therefore investigate whether the linguistic content of deepfakes affects attribution performance. In the baseline setup, support and query sets were disjoint in terms of text prompts used for audio generation.
In this experiment, prompts are randomly assigned to support or query sets, independent of the audio generator, so the same prompt may appear in support for one checkpoint and query for another. Using three random seeds, results are averaged in Table~\ref{tbl:baseline} (row~$5$). Checkpoint attribution shows a substantial performance drop, while architecture attribution is only slightly affected. Analysis of checkpoint-level confusions (not shown due to space limitation) indicates that VITS maintains strong checkpoint discrimination, whereas the other three architectures exhibit earlier iteration checkpoint confusion, suggesting that linguistic content definitely and significantly influences attribution.



\textbf{Dependency on vocoder.} Vocoders in TTS systems generate the final waveform, either as a stand-alone network (FastPitch, Grad-TTS, Matcha-TTS) or internally (VITS). Variations in vocoder architectures may lead SSL-derived features to capture vocoder characteristics rather than the generator itself, as prior studies have shown their strong impact on deepfake detection and attribution~\cite{10448016,10446331}. To examine this, we evaluate the attribution when all architectures use the same vocoder checkpoint, both in query and support. We report  averaged results across the three vocoder checkpoints (introduced in Section~\ref{sec:meth}) in row~6 of  Table~\ref{tbl:baseline}. VITS data remains in both support and query sets for cross-reference with the baseline. We notice only a small drop in performance, meaning that the TTS architectures maintain part of their own signatures and are captured by both SSL models. We also explore what happens if we change only the vocoder for a TTS architecture. For row~7 in Table~\ref{tbl:baseline} we maintained the default vocoders for the support set, and used the audio samples generated with the vocoders from the other architectures in the query set. The attribution performance is seriously affected in this case. No clear patterns of misattribution were observed in the confusion matrix, making these preliminary results merit further investigation.

\textbf{Dependency on speaker.} A third and most variable source of perturbation in TTS outputs is speaker identity. Commercial systems either provide multi-speaker models, conditioned on a speaker ID but with fixed weights across speakers, or allow voice cloning. Zero- or one-shot cloning uses reference audio without changing model weights, whereas advanced cloning finetunes the core model based on user-provided samples, updating the weights. In previous work~\cite{stan_interspeech25}, we showed that multi-speaker model attribution is not solely driven by speaker identity.

In this study, we replicate the finetuning approach by adapting pretrained model weights using a new speaker’s data for all four architectures. Results are reported in Table~\ref{tbl:baseline} (rows~8–10). In-domain splits (row~8) yield near-perfect attribution, but accuracy drops substantially when only the pretrained checkpoint (row~9) or pretrained plus intermediate checkpoints (row~10) are available. \bert appears more robust to speaker identity, likely due to its larger pretraining dataset capturing greater speaker variability (4.5M hours vs 450k for \xls). Notably, Grad-TTS samples are frequently misattributed to Matcha-TTS (65\%), and finetuned samples tend to align with low-iteration checkpoints, suggesting short adaptation (10k iterations) produces outputs similar to early LJSpeech training. These findings merit further study, to be addressed in future work.

\textbf{Zero-init checkpoints.} Attribution appears accurate and consistent across scenarios, prompting investigation of whether architectures themselves introduce inherent biases or artefacts. We use three random initializations (zero-init checkpoints) per architecture with no training, differing only in random seed, and generate 600 prompts. The outputs are essentially noise. In-domain attribution among these checkpoints (row~11, Table~\ref{tbl:baseline}) achieves perfect architecture-level ($F1=1.0$), but using them to attribute generated samples at different training stages, unfortunately, shows no correlation (row~12, $F1=0.1$ for both SSL models). Here, we could argue that the SSL models were not trained on such noisy data, and that passing these samples through the network will inevitably lead to unexpected results. Yet, we considered it to be an interesting setup worth exploring.



\vspace{-.2cm}
\section{Discussions and conclusions}
\label{sec:conc}

As the audio deepfake landscape grows more complex, accurate detection and attribution systems are essential to counteract potential misuse in fraud and deception. Some of the most effective systems rely heavily on features derived from large pretrained SSL models. However, the precise properties captured by these features and their sensitivity to minor variations in an audio generator’s weights remain unclear. In this work, we analyzed variations across the axes of text prompts, vocoders, and speaker identity, revealing that the two SSL models under study exhibit different error patterns. Overall, \bert demonstrates superior performance in out-of-domain scenarios, but is less effective in-domain; it also appears more influenced by textual content and less affected by speaker identity. Mismatches between pairs of TTS architectures and vocoders drastically affect both models. 

Nevertheless, this study requires further investigation of aspects such as zero-shot voice cloning models, multiple implementations of the same architecture, a broader set of speakers for adaptation, and a larger range of vocoder classes. These objectives are part of our ongoing work and will be addressed in future studies.

\clearpage

\textbf{Acknowledgements.} 
This work was co-funded by EU Horizon project AI4TRUST (No. 101070190), and by the Romanian Ministry of Research, Innovation and Digitization project DLT-AI SECSPP (id: PN-IV-P6-6.3-SOL-2024-2-0312).
Parts of this paper were rephrased using ChatGPT's free tier models.

\bibliographystyle{IEEEbib}
\bibliography{main}

\end{document}